\documentstyle[12pt]{article}
\newfont{\ffont}{msym10}                        
\newcommand{\beq}{\begin{equation}}             
\newcommand{\eeq}{\end{equation}}               
\newcommand{\bqry}{\begin{eqnarray}}            
\newcommand{\eqry}{\end{eqnarray}}              
\newcommand{\bqryn}{\begin{eqnarray*}}          
\newcommand{\eqryn}{\end{eqnarray*}}            
\newcommand{\NL}{\nonumber \\}                  
\newcommand{\preprint}[1]{\begin{table}[t]      
            \begin{flushright}                  
            \begin{large}{#1}\end{large}        
            \end{flushright}                    
            \end{table}}                        
\newcommand{\PD}[2]                             
    {\frac{\partial^{#2}}{\partial #1^{#2}}}    
\begin{document}
\preprint{TAUP-2115-93 \\  }
\title{Equilibrium Relativistic Mass Distribution
for Indistinguishable Events}
\author{\\ L. Burakovsky\thanks {Bitnet:BURAKOV@TAUNIVM.} \
and L.P. Horwitz\thanks
  {Bitnet:HORWITZ@TAUNIVM. Also at Department of Physics, Bar-Ilan
University, Ramat-Gan, Israel  } \\ \ }
\date{School of Physics and Astronomy \\ Raymond and Beverly Sackler
Faculty of Exact Sciences \\ Tel-Aviv University,
Tel-Aviv 69978, ISRAEL}
\maketitle
\begin{abstract}
A manifestly covariant relativistic statistical mechanics of the system
of $N$ indistinguishable events with motion in space-time parametrized
by an invariant ``historical time'' $\tau $ is considered. The
relativistic mass distribution for such a system is obtained from
the equilibrium solution of the generalized relativistic
Boltzmann equation by integration
over angular and hyperbolic angular variables. All the characteristic
averages are calculated. Expressions for the pressure and the density of
events are found and the relativistic equation of state is obtained. The
Galilean limit is considered; the theory is shown to pass over to the
usual nonrelativistic statistical mechanics of indistinguishable
particles.
\end{abstract}
\bigskip
{\it Key words:} special relativity, relativistic
Bose-Einstein/Fermi-Dirac, mass distribution

PACS: 03.30.+p, 05.20.Gg, 05.30.Ch, 98.20.--d
\bigskip
\section{Introduction}

With this paper we continue a series of works on manifestly covariant
relativistic statistical mechanics. In a
previous paper \cite{BH3} a manifestly covariant generalized
relativistic Boltzmann equation for both cases of the system of $N$
distinguishable and indistinguishable events treated simultaneously has
been made in the framework of a manifestly covariant classical and
quantum mechanics \cite{rqm}. These events,
considered as the fundamental dynamical objects of the theory, move in an
$8N$-dimensional phase space. Their motion is parametrized by a
continuous Poincar\'{e} invariant parameter $\tau $ called the
``historical time''.
The collection of events (called ``concatenation'' \cite {conc})
along each world line corresponds to a $particle$ in the usual sense;
e.g., the Maxwell conserved current is an integral over the history of
the charged event \cite{Jack}. Hence the evolution of the state of the
$N$-event system describes the history in
space and time of an $N$-particle system.

A generalized relativistic Boltzmann
equation was used to prove the $H$-theorem for evolution in $\tau$. In
the equilibrium limit the covariant forms of the known nonrelativistic
statistical distributions were obtained. Since these distributions are
the distributions of the 4-momenta of the events, $m^2=-p^2=-p^{\mu }
p_{\mu }$ is a random variable in a relativistic ensemble.

The case of distinguishable events was treated in a series of papers
\cite{HorShaSch},\cite{BH1},\cite{BH2}.
In [5] the simplest case of a narrow mass
shell $p^{2}=-m^{2}\cong -M^{2}$ was considered, where $M$ is a
given fixed parameter with the dimension of mass (an intrinsic property
of the particles), assumed to be the same for all the particles of the
system. The results obtained in this approximation are in agreement with
the well-known results of Synge \cite{Syn}
from an on-mass-shell relativistic kinetic theory.

In [6] the general case $0\leq m<\infty $ was considered and a
relativistic mass distribution was obtained. It has been shown
[7] that in the Galilean limit the
statistical mechanics of distinguishable events considered in [6] passes
over to the usual nonrelativistic statistical
mechanics of distinguishable particles.

In the present paper we consider the case of indistinguishable events.
We obtain a relativistic mass distribution for this case, from the
equilibrium solution of the manifestly covariant generalized Boltzmann
equation obtained in [1], by integration over angular and hyperbolic
angular variables. We calculate all the characteristic averages, obtain
expressions for the energy-momentum tensor, pressure and density of the
events, and derive the relativistic equation of state.

We show that in the Galilean limit the theory which we discuss here
passes over to the usual nonrelativistic statistical mechanics of
indistinguishable particles.

Finally, the criteria of validity of the theory presented in this paper
are obtained.

\section{Relativistic mass distribution}

We begin with the equilibrium relativistic distribution function obtained
in [1] (we use the metric $g^{\mu \nu }=(-,+,+,+)$ and $q\equiv q^\mu ,\;
p\equiv p^\mu $),
\beq
f_0(q,p)=C(q)\frac{1}{e^{-A(q)(p+p_c)^2-B(q)}\mp 1}\equiv
C(q)\frac{e^{A(q)(p+p_c)^2+B(q)}}{1\mp e^{A(q)(p+p_c)^2+B(q)}},\;A(q)>0.
\eeq

The function (1) must be normalized, according to [1], as
\beq
n(q)=\int f_{0}(q,p)d^4p=C(q)\int d^4p\frac{e^{A(p+p_c)^2+B}}{1\mp
e^{A(p+p_c)^2+B}},
\eeq
where $n(q)$ is the total number of events per unit space-time volume
in the system in the neighborhood of the point $q$. By expanding the
denominator in the integrand into a power series and introducing
hyperbolic variables $$\Omega ^4: m\geq 0,\;0\leq \theta \leq \pi ,\;
0\leq \varphi <2\pi ,\;-\infty <\beta <\infty ,$$
we can rewrite (2) as follows:
\bqry
n(q) & = & C(q)\sum _{n=1}^\infty (\pm 1)^{n+1}\int _{\Omega ^4}d^4p
e^{nA(p+p_c)^2+nB} \NL  & = & C(q)\sum _{n=1}^\infty (\pm 1)^{n+1}
\int _{\Omega ^4}m^3\sinh^2\beta \sin\theta dmd\beta d\theta d\varphi
\NL  &   & \times e^{-nAm^2-2nAm_cm\cosh \beta -n(Am_c^2-B)}.
\eqry
After some
calculations \cite{GrRy} we obtain the normalization relation
\beq
n(q)=C(q)\frac {\pi }{2A^2}\sum _{n=1}^\infty \frac{(\pm 1)^{n+1}}{n^2}
e^{-n(Am_c^2-B)}\Psi (2,2;Am^2_c).
\eeq
Here $\Psi (a,b;z)$ is the confluent hypergeometrical function of $z$
(ref.[10], p.257, sec.6.6).
After integration over angular and hyperbolic
angular variables we obtain from (3) the expression
\beq
n(q)=C(q)\frac {2\pi }{Am_c}\sum _{n=1}^\infty \frac{(\pm 1)^{n+1}}{n}
\int _{0}^{\infty }dmm^2e^{-n(Am_c^2-B)}e^{-nAm^2}K_1 (2nAm_cm),
\eeq
from which we identify the mass distribution function
\beq
f(m)=D\sum _{n=1}^\infty \frac{(\pm 1)^{n+1}}{n}m^2e^{-n(Am_c^2-B)}
e^{-nAm^2}K_1(2nAm_cm),
\eeq
where
\beq
D^{-1}=\frac{m_c}{4A}\sum _{n=1}^\infty \frac{(\pm 1)^{n+1}}{n^2}e^{-n(A
m_c^2-B)}\Psi (2,2;nAm_c^{2}),
\eeq
so that $f(m)$ is normalized according to
\beq
\int ^{\infty }_0 f(m)dm=1.
\eeq
In (5) and (6) $K_1$ is the Bessel function of the third kind
(imaginary argument), where, in general, $$K_{\nu }(z)=\frac{\pi i}{2}
e^{\pi i\nu /2}H_{\nu }^{(1)}(iz).$$
With the help of this distribution one can obtain the local average
value of an arbitrary function of mass $\phi (m)$:
\beq
\langle \phi (m)\rangle _q=\int _0^\infty \phi (m)f(m)dm.
\eeq
Let us obtain, for example, the local average values of mass and
mass squared in relativistic gas which represent the first two moments
of the distribution (6) (factors $m^\ell $ result in a change in the
first argument of the confluent hypergeometric function and its
normalization):
\beq
\langle m\rangle _q=\frac{3\pi }{8A^{\frac{1}{2}}}\frac{\sum _{n=1}^
\infty \{(\pm 1)^{n+1}e^{-n(Am_c^2-B)}\Psi (\frac{5}{2},2;nAm_c^2)/
n^{\frac{5}{2}}\}}{\sum _{n=1}^\infty \{(\pm 1)^{n+1}e^{-n(Am_c^2-B)}
\Psi (2,2;nAm_c^2)/n^2\}},
\eeq
\beq
\langle m^2\rangle _q=\frac{2}{A}\frac{\sum _{n=1}^\infty \{(\pm 1)^
{n+1}e^{-n(Am_c^2-B)}\Psi (3,2;nAm_c^2)/n^3\}}{\sum _{n=1}^
\infty \{(\pm 1)^{n+1}e^{-n(Am_c^2-B)}\Psi (2,2;nAm_c^2)/n^2\}}.
\eeq
More generally,
\beq
\langle m^\ell \rangle _q=\Gamma (\frac{\ell }{2}+1)\Gamma (\frac{\ell }
{2}+2)A^{-\frac{\ell }{2}}\frac{\sum _{n=1}^\infty \{(\pm 1)^{n+1}
e^{-n(Am_c^2-B)}\Psi (2+\frac{\ell }{2},2;nAm_c^2)/n^{2+\frac{\ell }{2}}
\}}{\sum _{n=1}^\infty \{(\pm 1)^{n+1}e^{-n(Am_c^2-B)}\Psi (2,2;nAm_c^2)
/n^2\}}.
\eeq
Now, as in [5],[6], we define absolute temperature through the relation
\beq
2Am_c=\frac {1}{k_BT},
\eeq
which implies that in thermal equilibrium $Am_c$ is independent of $q$.
Hence
\beq
\langle m\rangle =\frac {3\pi }{8}\sqrt {2m_ck_BT}
\frac{\sum _{n=1}^\infty \{(\pm 1)^{n+1}e^{-n(\frac{m_c}{2k
_BT}-B)}\Psi (\frac{5}{2},2;n\frac{m_c}{2k_BT})/n^{\frac{5}{2}}\}}
{\sum _{n=1}^\infty \{(\pm 1)^{n+1}e^{-n(\frac{m_c}{2k_BT}-B)}
\Psi (2,2;n\frac{m_c}{2k_BT})/n^2\}}.
\eeq
In the limit $T\rightarrow 0$ it follows from the asymptotic
formula for $z\rightarrow \infty $ \cite{conf2}
\beq
\Psi (a,b;z)\sim z^{-a}[1+\sum _{n=1}^\infty (-1)^n
\frac {a(a+1)\cdots (a+n-1)(1+a-b)(2+a-b)\cdots (n+a-b)}{n!z^n}]
\eeq
that
\beq
\langle m\rangle \cong \frac{3\pi }{4}k_BT\frac{\sum _{n=1}^\infty \{
(\pm 1)^{n+1}e^{-n(\frac{m_c}{2k_BT}-B)}/n^5\}}{\sum _{n=1}^\infty
\{(\pm 1)^{n+1}e^{-n(\frac{m_c}{2k_BT}-B)}/n^4\}}.
\eeq
One can also obtain in this limit
\beq
\langle m^2\rangle \cong 8(k_BT)^2\frac{\sum _{n=1}^\infty \{
(\pm 1)^{n+1}e^{-n(\frac{m_c}{2k_BT}-B)}/n^6\}}{\sum _{n=1}^\infty
\{(\pm 1)^{n+1}e^{-n(\frac{m_c}{2k_BT}-B)}/n^4\}}.
\eeq

Let us now calculate the first two moments of the distribution (1):
\beq
\langle p^{\mu }\rangle _q=\frac{\int d^4pp^\mu f_0(q,p)}{\int d^4pf_0(q,
p)}=\frac {\int d^4pp^\mu \frac{e^{A(p+p_c)^2+B}}{1\mp e^{A(p+p_c)^2+B}}}
{\int d^4p\frac{e^{A(p+p_c)^2+B}}{1\mp e^{A(p+p_c)^2+B}}}=\frac{\sum _{n=
1}^{\infty }(\pm 1)^{n+1}\int d^4pp^\mu e^{nA(p+p_c)^2+nB}}{\sum _{n=
1}^{\infty }(\pm 1)^{n+1}\int d^4pe^{nA(p+p_c)^2+nB}},
\eeq
\beq
\langle p^{\mu }p^\nu \rangle _q=\frac{\int d^4pp^\mu p^\nu f_0(q,p)}{
\int d^4pf_0(q,p)}=\frac{\sum _{n=1}^{\infty }(\pm 1)^{n+1}\int d^4p
p^\mu p^\nu e^{nA(p+p_c)^2+nB}}{\sum _{n=1}^{\infty }(\pm 1)^{n+1}
\int d^4pe^{nA(p+p_c)^2+nB}}.
\eeq
Define the following functions:
\bqry
F_0 & = & \sum _{n=1}^{\infty }(\pm 1)^{n+1}\int d^4pe^{nA(p+p_c)^2+
nB}\equiv \int d^4pf_0(q,p), \\
F_1 & = & \sum _{n=1}^
\infty \frac{(\pm 1)^{n+1}}{n}\int d^4pe^{nA(p+p_c)^2+nB}, \\
F_2 & = & \sum _{n=1}^\infty
\frac{(\pm 1)^{n+1}}{n^2}\int d^4pe^{nA(p+p_c)^2+nB}.
\eqry
We have
\bqry
\langle p^{\mu }\rangle _q & = & F_0^{-1}(\frac{1}{2A}\frac{\partial F_1}
{\partial p_{c\mu }}-p^\mu _c), \\ \langle p^\mu p^\nu \rangle _q & = &
F_0^{-1}\left(\frac{1}{4A^2}\frac{\partial ^2F_2}{\partial p_{c\mu }
\partial p_{c\nu }}-\langle p^\mu \rangle p_c^\nu -\langle p^\nu
\rangle p_c^\mu -p_c^\mu p_c^\nu \right).
\eqry
Calculating $F_1$ and $F_2$ in explicit form and using the relations
\bqryn
\frac{\partial }{\partial p_{c\mu }} & = & -2p_c^\mu \frac{\partial }
{\partial m_c^2}, \\ \frac{\partial ^2}{\partial p_{c\mu }\partial p_{c
\nu }} & = & 4p_c^\mu p_c^\nu \frac{\partial ^2}{\partial (m_c^2)^2}-2g^{
\mu \nu }\frac{\partial }{\partial m_c^2},
\eqryn
we obtain with the help of the formula (ref.[10], p.257,
sec.6.6) $$\frac{d}{dz}\Psi (a,b;z)=-a\Psi (a+1,b+1;z)$$
\bqry
\langle p^{\mu }\rangle _q & = & 2p_c^{\mu }\frac{\sum _{n=1}^\infty
\{(\pm 1)^{n+1}e^{-n(Am_c^2-B)}\Psi (3,3;nAm_c^2)/n^2\}}{\sum _{n=1}
^\infty \{(\pm 1)^{n+1}e^{-n(Am_c^2-B)}\Psi (2,2;nAm_c^2)/n^2\}}, \\
\langle p^\mu p^\nu \rangle _q & = & 6p_c^\mu p_c^\nu \frac{\sum _{n=1}^
\infty \{(\pm 1)^{n+1}e^{-n(Am_c^2-B)}\Psi (4,4;nAm_c^2)/n^2\}}
{\sum _{n=1}^\infty \{(\pm 1)^{n+1}e^{-n(Am_c^2-B)}\Psi (2,2;nAm_c^2)/
n^2\}} \NL  &  & +\frac{g^{\mu \nu }}{A}\frac{\sum _{n=1}^\infty
\{(\pm 1)^{n+1}e^{-n(Am_c^2-B)}\Psi (3,3;nAm_c^2)/n^3\}}{\sum _{n=1}
^\infty \{(\pm 1)^{n+1}e^{-n(Am_c^2-B)}\Psi (2,2;nAm_c^2)/n^2\}}.
\eqry
As in [5],[6], we make a Lorentz transformation to the local average
motion rest frame moving with the relative velocity
$${\bf u}=\frac{{\bf p}_c}{E_c},$$ in order to obtain the local
energy density.

The rest frame energy is $$\langle E^{\prime }\rangle _q=\frac
{\langle E\rangle _q-{\bf u}\cdot {\bf p}}{\sqrt {1-{\bf u}^2}},$$
so that
\beq
\langle E^{\prime }\rangle _q=2m_c\frac{\sum _{n=1}^\infty
\{(\pm 1)^{n+1}e^{-n(Am_c^2-B)}\Psi (3,3;nAm_c^2)/n^2\}}{\sum _{n=1}
^\infty \{(\pm 1)^{n+1}e^{-n(Am_c^2-B)}\Psi (2,2;nAm_c^2)/n^2\}}.
\eeq

To obtain the pressure and the density of events in our ensemble,
as in [5],[6], we study the $particle$ energy-momentum
tensor defined by the $R^4$ density
\beq
T^{\mu \nu }(q)=\sum _{i}\int d\tau \frac {p^{\mu }_i p^{\nu }_i }{M}
\delta ^4 (q-q_i(\tau )).
\eeq
Using the result of [5]
\beq
\langle T^{\mu \nu }(q)\rangle _q =T_{\triangle V}\int d^4 pf_0 (q,p)
\frac {p^\mu p^\nu }{M}
\eeq
and the expression (26) for $\langle p^\mu p^\nu \rangle _q $, we obtain
$$\langle T^{\mu \nu }(q)\rangle _q =\frac {T_{\triangle V}n(q)}{M}\left[
\frac{g^{\mu \nu }}{A}\frac{\sum _{n=1}^\infty \{(\pm 1)^{n+1}
e^{-n(Am_c^2-B)}\Psi (3,3;nAm_c^2)/n^3\}}{\sum _{n=1}
^\infty \{(\pm 1)^{n+1}e^{-n(Am_c^2-B)}\Psi (2,2;nAm_c^2)/n^2\}}\right.$$
\beq
\left. +6p_c^\mu p_c^\nu \frac{\sum _{n=1}^\infty
\{(\pm 1)^{n+1}e^{-n(Am_c^2-B)}\Psi (4,4;nAm_c^2)/n^2\}}{\sum _{n=1}
^\infty \{(\pm 1)^{n+1}e^{-n(Am_c^2-B)}\Psi (2,2;nAm_c^2)/n^2\}}\right].
\eeq
In this expression $T_{\triangle V}$ is the average passage interval in
$\tau $ for the events which pass through the small four-volume
$\triangle V $ over the point $q$ of $R^4$.

The formula for the stress-energy tensor of a perfect fluid has the form
[5,(3.39)]
\beq
\langle T^{\mu \nu }(q)\rangle _q =pg^{\mu \nu }-(p+\rho )\frac {\langle
 p^{\mu }\rangle _q \langle p^{\nu }\rangle _q }
{\langle p^{\lambda }\rangle _q \langle p_{\lambda }\rangle _q},
\eeq
where $p$ is the pressure and $\rho $ is the density of energy at $q$.

According to  (25),
\beq
\frac{\langle p^{\mu }\rangle _q }
{\sqrt{-\langle p^{\lambda }\rangle _q \langle p_{\lambda }\rangle _q }}
=\frac{p^{\mu }_c}{m_c},
\eeq
hence
\beq
p=\frac{T_{\triangle V}n(q)}{AM}\frac{\sum _{n=1}^\infty \{(\pm 1)^{n+1}
e^{-n(Am_c^2-B)}\Psi (3,3;nAm_c^2)/n^3\}}{\sum _{n=1}
^\infty \{(\pm 1)^{n+1}e^{-n(Am_c^2-B)}\Psi (2,2;nAm_c^2)/n^2\}}.
\eeq
and
\beq
p+\rho =\frac{6T_{\triangle V}n(q)m^2_c}{M}\frac{\sum _{n=1}^\infty
\{(\pm 1)^{n+1}e^{-n(Am_c^2-B)}\Psi (4,4;nAm_c^2)/n^2\}}{\sum _{n=1}
^\infty \{(\pm 1)^{n+1}e^{-n(Am_c^2-B)}\Psi (2,2;nAm_c^2)/n^2\}}.
\eeq
To interpret these results, as in [5],[6], we should
calculate the average (conserved) $particle$
four-current, which has the microscopic form
\beq
J^{\mu }(q)=\sum _i \int \frac {p^{\mu }_i}{M}\delta ^4 (q-q_i(\tau ))d
\tau .
\eeq
Using the result of [5]
\beq
\langle J^{\mu }(q)\rangle _q =T_{\triangle V}\int d^4 p\frac{p^{\mu }}
{M}f_0 (q,p)
\eeq
and expression (25) for $\langle p^{\mu }\rangle _q$, we obtain
\beq
\langle J^{\mu }(q)\rangle _q =\frac{2T_{\triangle V}n(q)}{M}p^{\mu }_c
\frac{\sum _{n=1}^\infty \{(\pm 1)^{n+1}
e^{-n(Am_c^2-B)}\Psi (3,3;nAm_c^2)/n^2\}}{\sum _{n=1}
^\infty \{(\pm 1)^{n+1}e^{-n(Am_c^2-B)}\Psi (2,2;nAm_c^2)/n^2\}}.
\eeq
In the local rest frame $p^{\mu }_c =(m_c,{\bf 0})$,
\beq
\langle J^0(q)\rangle _q=\frac{2T_{\triangle V}n(q)m_c}{M}\frac{\sum _{n=
1}^\infty \{(\pm 1)^{n+1}e^{-n(Am_c^2-B)}\Psi (3,3;nAm_c^2)/n^2\}}{\sum _
{n=1}^\infty \{(\pm 1)^{n+1}e^{-n(Am_c^2-B)}\Psi (2,2;nAm_c^2)/n^2\}}.
\eeq
Defining the density of $paricles$ per unit space volume as
\beq
N_{0}(q)=\langle J^{0}(q)\rangle _q,
\eeq
we obtain the relativistic equation of state
$$p=\frac{N_0}{2Am_c}\frac{\sum _{n=1}^\infty \{(\pm 1)^{n+1}
e^{-n(Am_c^2-B)}\Psi (3,3;nAm_c^2)/n^3\}}{\sum _{n=1}
^\infty \{(\pm 1)^{n+1}e^{-n(Am_c^2-B)}\Psi (3,3;nAm_c^2)/n^2\}}$$
\beq
=N_0k_BT\frac{\sum _{n=1}^\infty \{(\pm 1)^{n+1}e^{-n(\frac{m_c}{2k_BT}
-B)}\Psi (3,3;n\frac{m_c}{2k_BT})/n^3\}}{\sum _{n=1}^\infty \{(\pm 1)^
{n+1}e^{-n(\frac{m_c}{2k_BT}-B)}\Psi (3,3;n\frac{m_c}{2k_BT})/n^2\}}.
\eeq

In limiting case $T\rightarrow 0$ the general expression for the
distribution function (6) can be simplified. Indeed, it is seen from (17)
and (27) that in this case $\langle m^2\rangle \sim (k_BT)^2\ll \langle
E^{\prime}\rangle m_c\sim m_c(k_BT).$ Therefore, one can
neglect $m^2=-p^2$ in comparison with $2App_c$ in
the exponent of the initial distribution (1); hence, we begin with
\beq
f_0^{\prime }(p,q)\cong C(q)\frac{1}{e^{Am_c^2-2App_c-B}\mp 1},
\eeq
and after integration and normalization obtain
\beq
f^{0}(m)=D^0\sum _{n=1}^\infty \frac{(\pm 1)^{n+1}}{n}m^2e^{-n(Am_c^2-B)}
K_1(2nAm_cm),
\eeq
\beq
(D^0)^{-1}=\frac{2}{(2Am_c)^3}\sum _{n=1}^\infty \frac{(\pm 1)^{n+1}}{n^4
}e^{-n(Am_c^2-B)}.
\eeq
The distribution function (42) gives $$\langle m^{\ell }\rangle ^{0}=
\Gamma (\frac{\ell }{2}+1)\Gamma (\frac{\ell }{2}+2)(Am_c)^{-\ell }\frac
{\sum _{n=1}^\infty \{(\pm 1)^{n+1}e^{-n(Am_c^2-B)}/n^{4+\ell }\}}
{\sum _{n=1}^\infty \{(\pm 1)^{n+1}e^{-n(Am_c^2-B)}/n^4\}}$$
\beq
=\Gamma (\frac{\ell }{2}+1)\Gamma (\frac{\ell }{2}+2)(2k_BT)^{\ell }\frac
{\sum _{n=1}^\infty \{(\pm 1)^{n+1}e^{-n(\frac{m_c}{2k_BT}-B)}/n^{4+\ell
}\}}{\sum _{n=1}^\infty \{(\pm 1)^{n+1}e^{-n(\frac{m_c}{2k_BT}-B)}/n^4\}
},
\eeq
which coincides with the low-temperature limit of (12).
Similarly, one can obtain in this case
\beq
\langle p^{\mu }\rangle ^{0}=\frac{p_c^{\mu }}{m_c}4k_BT\frac{\sum _{n=1}
^\infty \{(\pm 1)^{n+1}e^{-n(\frac{m_c}{2k_BT}-B)}/n^5\}}{\sum _{n=1}^
\infty \{(\pm 1)^{n+1}e^{-n(\frac{m_c}{2k_BT}-B)}/n^4\}},
\eeq
which coincides with the low-temperature limit of (25).

To interpret these results, consider the initial equilibrium
distribution function $f_0(q,p)$ in the local rest frame $p^\mu _c=(m_c,
{\bf 0}).$ Using (13) one obtains $$f_0(q,p)=C(q)\frac{1}
{e^{-A(p+p_c)^2-B}\mp 1}=C(q)\frac{1}{e^{Am^2+2Am_cE+(Am_c^2-B)}\mp 1}$$
\beq
=C(q)\frac{1}{\exp \{\frac{1}{k_BT}
[E+\frac{m^2}{2m_c}+(\frac{m_c}{2}-k_BT\cdot B)]\}\mp 1}.
\eeq
Therefore, one can put
\beq
m_c\equiv \frac{M}{\mu _K},
\eeq
\beq
B\equiv \frac{1}{k_BT}(\mu +\frac{m_c}{2}),
\eeq
and hence
to identify (46) with the grand canonical distribution function $$
\frac{1}{\exp \{\frac{1}{k_BT}[E-\mu +\mu _K\frac{m^2}{2M}]\}\mp 1},$$
obtained previously in \cite{HSP} for the static Gibbs ensembles. In the
latter formula $\mu $ and $\mu _K$ are relativistic chemical and mass
potentials, respectively [11].
Hence $Am_c^2-B\equiv -\mu /k_BT,$ and $\exp \{-n(Am_c^2-B)\}$ should be
identified with $\exp (\frac{n\mu }{k_BT})$ in all formulas
containing this exponent, throughout the paper.

For example, the relativistic equation of state (40) now reads
\beq
p=N_0k_BT\frac{\sum _{n=1}^\infty \{(\pm 1)^{n+1}e^{\frac{n\mu }{k_BT}}
\Psi (3,3;\frac{nM}{2\mu _Kk_BT})/n^3\}}{\sum _{n=1}^\infty \{(\pm 1)^
{n+1}e^{\frac{n\mu }{k_BT}}\Psi (3,3;\frac{nM}{2\mu _Kk_BT})/n^2\}}.
\eeq
Since at thermal equilibrium there is no dependence on $q,$ according to
[1], instead of $f_0(q,p)$ one can use equilibrium relativistic
distribution function\footnote{Clearly, all the results of the paper,
since obtained for thermal equilibrium, remain valid.}
\beq
f_0(p)=\int d^4qf_0(q,p)=\frac{1}{e^{(E-\mu +\mu _K\frac{m^2}{2M})/k_BT}
\mp 1},
\eeq
normalized as
\beq
\int \frac{d^4p}{(2\pi )^4}f_0(p)=n,
\eeq
where $n\equiv \frac{N}{V^{(4)}}$ is the event number density.

\section{Galilean limit of relativistic mass distribution}

The Galilean limit of manifestly covariant relativistic theory which we
discuss here has been considered in a series of papers
[7],[11],\cite{HR}, by taking $c\rightarrow \infty$ (compared to
all other velocities). It has been shown [7] that in the Galilean
limit the relativistic relation between the energy $E$ and the mass $m$
\beq
E^2=m^2+{\bf p}^2
\eeq
transforms to
\beq
E=m+\frac{{\bf p}^2}{2M},
\eeq
where the Galilean mass
$M$ coincides with the particle's intrinsic parameter.
At the same time the quantity
\beq
\eta =c^2(m-M)
\eeq
may take any value, however, finite, as $c
\rightarrow \infty ;$ or, equivalently, \\ $m=M(1+O(1/c^2)).$

In case of an equilibrium relativistic ensemble of distinguishable events
such a transformation of the relativistic relation between $E$ and $m$
gives rise to the usual nonrelativistic Maxwell-Boltzmann distribution
of ${\bf p}^2/2M.$

Now we shall consider the case of indistinguishable events. We start with
the low-temperature form (41) of initial equilibrium relativistic
distribution (1) which is normalized as follows,
\beq
n(q)=C(q)\int d^4p\frac{1}{e^{Am_c^2-2App_c-B}\mp 1}.
\eeq
This integral written in the local rest frame takes the form
\beq
\int dEd^3{\bf p}\sum _{n=1}^\infty (\pm 1)^{n+1}e^{-2nAm_cE}e^{-n(Am_c^2
-B)}\equiv \int dEd^3{\bf p}\sum _{n=1}^\infty (\pm 1)^{n+1}e^{-\frac{nE}
{k_BT}}e^{\frac{n\mu }{k_BT}},
\eeq
or, in view of (53),(54) and the relation $-\triangle \leq \eta \leq
\triangle $ [7],
\beq
\sum _{n=1}^\infty \int _{M-\triangle }^{M+\triangle }
dme^{-\frac{nm}{k_BT}}\int d^3{\bf p}e^{-\frac{n}{k_BT}(\frac{{\bf p}^2}
{2M}-\mu )}.
\eeq
Integration on $m$ gives
\beq
\int _{M-\triangle }^{M+\triangle }dme^{-\frac{nm}{k_BT}}=
2\frac{k_BT}{n}e^{-\frac{nM}{k_BT}}\sinh \frac{n\triangle }{k_BT}.
\eeq
Since $\triangle $ may take any infinitesimal value\footnote{It
corresponds to infinitely sharp mass shell $\triangle =c^2\mid m-M
\mid [7].$}, but not zero (it enters the normalization factor), we obtain
in the limit $\triangle \rightarrow 0$ $$\frac{k_BT}{n}\sinh \frac
{n\triangle }{k_BT}\cong \triangle ,$$ and therefore
\beq
n(q)=2\triangle
C(q)\int d^3{\bf p}\sum _{n=1}^\infty (\pm 1)^{n+1}e^{-\frac
{n}{k_BT}(\frac{{\bf p}^2}{2M}-\mu +M)}\equiv 2\triangle
C(q)\int d^3{\bf p}\frac{1}{e^{(\frac{{\bf p}^2}{2M}-\tilde{\mu })
/k_BT}\mp 1},
\eeq
which is the usual (normalized) nonrelativistic
Bose-Einstein/Fermi-Dirac distribution for $e\equiv \frac{{\bf p}^2}{2M},
$
\beq
f(e)=\tilde{D}\frac{e^{1/2}}{\exp (\frac{e-\tilde{\mu }}{k_BT})\mp 1},
\eeq
\beq
\tilde{D}^{-1}=\Gamma \left(\frac{3}{2}\right)(k_BT)^{3/2}\sum _{n=1}^
\infty \frac{(\pm 1)^{n+1}}{n^{3/2}}e^{\frac{n\tilde{\mu }}{k_BT}}.
\eeq
In the latter formulas
\beq
\tilde{\mu }=\mu -M
\eeq
is the usual nonrelativistic chemical potential.

A precise equation for $\mu =\mu (n,\mu _K,T)$
can be obtained from (50),(51) by performing integration:
\beq
n=\frac{1}{(2\pi )^3}\frac{M^2}{\mu _K^2}(k_BT)^2\sum _{s=1}^\infty
\frac{(\pm 1)^{s+1}}{s^2}e^{\frac{s\mu }{k_BT}}
\Psi (2,2;\frac{sM}{2\mu _Kk_BT}).
\eeq
The corresponding equation for the low-temperature case is obtained from
(41),(51), or directly from (63), using the asymptotic formula (15),
\beq
n=\frac{(k_BT)^4}{2\pi ^3}\sum _{s=1}^\infty \frac{(\pm 1)^{s+1}}{s^4}
e^{\frac{s\mu }{k_BT}}.
\eeq
By means of the formula
\beq
\sum _{s=1}^\infty \frac{(\pm 1)^{s+1}}{s^p}e^{\frac{s\mu }{k_BT}}
\equiv \sum _{s=1}^\infty \frac{(\pm 1)^{s+1}}{s^p}z^{s}=\pm
\sum _{s=1}^\infty \frac{(\pm z)^{s}}{s^p}=\pm Li_p(\pm z),
\;\;\;z\equiv e^{\frac{\mu }{k_BT}},
\eeq
where\footnote{For $\mid z\mid \geq 1$ the function
$Li_\nu (z)$ is defined as the analytic continuation of this series.}
\beq
Li_\nu (z)\equiv \sum _{k=1}^\infty \frac{z^k}{k^\nu },\;\;\;\mid z\mid <
1\;\;\;\;{\rm or}\;\;\mid z\mid =1,\;\;Re \;\nu >1,
\eeq
is the so-called {\it polylogarithm}
\cite{Prud}, the equation (64) can be rewritten as follows,
\beq
n=\pm \frac{(k_BT)^4}{2\pi ^3}Li_4(\pm e^{\frac{\mu }{k_BT}}).
\eeq
For comparison, the equation of nonrelativistic statistical mechanics for
$\tilde{\mu }=\tilde{\mu }(N_0,T)$ is obtained from the normalization
relation of the distribution function (59),
\beq
\int \frac{d^3{\bf p}}{(2\pi )^3}\frac{1}{e^{(\frac{{\bf p}^2}{2M}-
\tilde{\mu })/k_BT}\mp 1}=N_0,
\eeq
$N_0$ being defined by (39), and is found to be
\beq
N_0=\pm \left(\frac{Mk_BT}{2\pi }\right)^{3/2}\!\!Li_{\frac{3}{2}}
(\pm e^{\frac{\tilde{\mu }}{k_BT}}).
\eeq

\section{Validity criteria}

Now we wish to consider the limits of applicability of the theory which
we discussed in the present paper and obtain explicit criteria for its
validity.

It is clear from the general considerations that the theory is valid at
medium and low temperatures\footnote{At high temperatures it goes over to
relativistic statistical mechanics of distinguishable events considered
previously in [6].} still far from absolute zero, at which the degeneracy
effects manifest themselves. The explicit criterions of its validity can
be obtained as follows:

1) Since in any system of macroscoping size the spacing between
successive momentum levels should be exceedingly small compared to the
characteristic thermal momentum $k_BT,$ one has\footnote{This
condition implies that in the thermodynamic limit,
summation over the discrete variables $n_i$ can be
replaced by integration over the continuous variable $p,$ where $n_i$
is the occupation number in the event energy-momentum space.}
\beq
\frac{p^\mu _n-p^\mu _{n-1}}{k_BT}=\frac{\hbar }{L^{(\mu )}k_BT}<<1,
\eeq
where $L^{(\mu )}$ is a
character size of a system in $\mu $-direction. Taking into account the
four similar relations for $\mu =0,1,2,3,$ (70) can be rewritten as
follows (in the system of units we use),
\beq
\frac{N}{L^4(k_BT)^4}<<1,\;\;\;{\rm or}\;\;\frac{n}{(k_BT)^4}<<1,
\eeq
where $L$ is a character size of a system, $L^4\sim V^{(4)},$ the total
four-volume, occupied by the system in four-dimensional space-time,
and $n=\frac{N}{V^{(4)}}$ is the event number density.
Hence, the criterion of validity is determined by a dimensionless  number
$\delta ,$ conventionally called the {\it degeneracy parameter:}
\beq
\delta =\frac{n}{(k_BT)^4}<<1.
\eeq
This criterion can be violated at low temperatures near absolute zero or
at high event densities. The two possible situations constitute the
so-called degenerate case and will be treated in the following research
[13].

2) In order that the expansion $$\frac{1}{e^{-A(p+p_c)^2-B}\mp 1}=\sum _{
n=1}^\infty (\pm 1)^{n+1}e^{nA(p+p_c)^2+nB},$$ which is basic for the
theory, be valid, we must have $$e^{A(p+p_c)^2+B}<1,$$ or,
equivalently, $$-A(p+p_c)^2-B>0.$$ This relation can be rewritten in the
local rest frame, using $\cosh \beta \geq 1,$ $$Am^2+2Amm_c\cosh \beta +
Am_c^2-B\geq A(m+m_c)^2-B>0.$$ Since the latter must be valid for any $m,
$ we deduce that $$B<Am_c^2\equiv \frac{m_c}{2k_BT}.$$ Using the relation
(48), $B\equiv \frac{m_c}{2k_BT}+\frac{\mu }{k_BT},$ we finally obtain
\beq
\mu <0.
\eeq
This criterion is analogous to the nonrelativistic one and holds at all
temperatures far from absolute zero, similar to the nonrelativistic case.

3) Considering $m_c$ as a parameter, in order to determine absolute
temperature for thermal equilibrium, one should have, in accordance with
(13), $2Am_c>0.$ It leads to $m_c>0,$ since $A>0,$ and,
through (47), to
\beq
\mu _K\geq 0.
\eeq
The three conditions (72)-(74) constitute the set of criteria for
validity of the theory discussed in the present paper.

The dependencies of $\mu $ and $\mu _K$ on $T$ are determined by equation
(63) which in the low-temperature limit transforms to (67).
The corresponding equation for the high-temperature case can be obtained
from the normalization relation of relativistic statistical mechanics of
distinguishable events discussed in [6], $$n= \frac{1}{(2\pi )^3}\frac{M^
2}{\mu _K^2}(k_BT)^2e^{\frac{\mu }{k_BT}}\Psi (2,2;\frac{M}{2\mu _Kk_BT}
),$$ by taking the limit $T\rightarrow \infty $ [6,(30)]:
\beq
n=\frac{1}{4\pi ^3}\frac{M}{\mu _K}(k_BT)^3e^{\frac{\mu }{k_BT}}.
\eeq
Summarizing, we can write down the following relations, taking into
account the three cases considered above, $$n=\left\{ \begin{array}{ll}
\pm \frac{1}{2\pi ^3}(k_BT)^4Li_4(\pm e^{\frac{\mu }{k_BT}}), & {\rm low
\;\;temperatures} \\ \frac{1}{(2\pi )^3}\frac{M^2}{\mu _K^2}(k_BT)^2\sum
_{s=1}^\infty \{(\pm 1)^{s+1}e^{\frac{s\mu }{k_BT}}\Psi (2,2;
\frac{sM}{2\mu _Kk_BT})/s^2\}, & {\rm intermediate\;\;case} \\
\frac{1}{4\pi ^3}\frac{M}{\mu _K}(k_BT)^3e^{\frac{\mu }{k_BT}}. &
{\rm high\;\;temperatures} \end{array} \right. $$

\section{Concluding remarks}

We have considered an equilibrium relativistic ensemble of
indistinguishable events, described by an equilibrium relativistic
distribution function representing covariant generalization of the known
nonrelativistic Bose-Einstein/Fermi-Dirac distributions, with variable
mass. We have shown that this distribution can be identified with a grand
canonical distribution function first introduced by Horwitz, Schieve and
Piron in their work on static Gibbs ensembles.

We have found that an equilibrium state of such a system is characterized
by a well-defined mass distribution, following directly from the
equilibrium relativistic distribution, by integration over angular and
hyprebolic angular variables. The normalization relation of this
distribution represents an equation linking the basic thermodynamic
properties of the system $\mu ,\mu _K,n,T.$

The dependence of $\mu $ on $T$ is explicitly defined only for low
temperatures. Although the dependence of $\mu _K$ on $T$ is not
determined, several remarks on this point can be made.
Miller and Suhonen \cite{MS}, studying possible applications of grand
canonical distribution function of ref. [11] in high-energy physics,
showed that at low temperatures one should have $\mu _K<<1,$ whenever at
high temperatures $\mu _K>>1.$ The former relation may also justify the
fact that it is possible to omit the term $\frac{\mu _Km^2}{2M}$ in
comparison with $(E-\mu )$ in the exponent of the equilibrium
relativistic distribution function in the low-temperature limit.

Finally, we have shown that in the Galilean limit a statistical mechanics
of indistinguishable events discussed in the present paper passes over to
the usual nonrelativistic statistical mechanics of indistinguishable
particles.

\end{document}